\documentclass[a4paper]{article}
\usepackage{hyperref}
\usepackage{color}
\usepackage{INTERSPEECH2020}

\title{Enhancing Speech Intelligibility in Text-To-Speech Synthesis using Speaking Style Conversion}
\name{Dipjyoti Paul$^1$, Muhammed PV Shifas$^1$, Yannis Pantazis$^2$ and Yannis Stylianou$^1$}
\address{
  $^1$Computer Science Department, University of Crete, Greece\\
  $^2$Inst. of Applied and Computational Mathematics, Foundation for Research and Technology - Hellas}
\email{\{dipjyotipaul,shifaspv,yannis\}@csd.uoc.gr, pantazis@iacm.forth.gr}

\begin{document}

\maketitle

\begin{abstract}

The increased adoption of digital assistants makes text-to-speech (TTS) synthesis systems an indispensable feature of modern mobile devices. It is hence desirable to build a system capable of generating highly intelligible speech in the presence of noise. Past studies have investigated style conversion in TTS synthesis, yet degraded synthesized quality often leads to worse intelligibility. To overcome such limitations, we proposed a novel transfer learning approach using Tacotron and WaveRNN based TTS synthesis. The proposed speech system exploits two modification strategies: (a) Lombard speaking style data and (b) Spectral Shaping and Dynamic Range Compression (SSDRC) which has been shown to provide high intelligibility gains by redistributing the signal energy on the time-frequency domain. We refer to this extension as Lombard-SSDRC TTS system. Intelligibility enhancement as quantified by the Intelligibility in Bits ($SIIB^{Gauss}$) measure shows that the proposed Lombard-SSDRC TTS system shows significant relative improvement between 110\% and 130\% in speech-shaped noise (SSN), and 47\% to 140\% in competing-speaker noise (CSN) against the state-of-the-art TTS approach. Additional subjective evaluation shows that Lombard-SSDRC TTS successfully increases the speech intelligibility with relative improvement of 455\% for SSN and 104\% for CSN in median keyword correction rate compared to the baseline TTS method.

\end{abstract}


\section{Introduction}

Humans often modify their speaking style to make the spoken message more comprehensible under challenging, noisy environments. Adapting to such speaking style is called the Lombard effect, and the resulting speech exhibits changes in both acoustic and phonetic properties such as an increase in vocal intensity, a decrease in spectral tilt, variations in formant frequencies as well as in phoneme duration \cite{lu2008speech, hansen1996analysis}.

Over the years, text-to-speech (TTS) systems have become more prevalent with a substantial range of applications including personal voice assistants, public address systems and navigation devices. In a quiet environment, the intelligibility of synthetic speech corresponds to that of natural speech. However, the intelligibility is typically fallen below the level of natural speech in noisy conditions \cite{cooke2013evaluating}. Listeners in real-world scenarios often hear speech in noisy surroundings where the intelligibility of synthetic speech is also compromised. Therefore, highly efficient TTS systems which are able to simulate Lombard effect
and make the speech more intelligible are essential for the end listeners. Such speaking style conversion
retains the linguistic and speaker-specific information of the original speech.

A considerable amount of research has been conducted on speaking style modification during the last years. Signal processing approaches such as cepstral modifications \cite{valentini2014intelligibility}, spectral shaping \cite{erro2014enhancing} and glimpse proportion measure with dynamic range compression \cite{valentini2013combining} were adopted to mimic the acoustic changes observed in the production of Lombard speech. Voice transformation techniques have been implemented to learn the mapping between normal 
speech and speech that is generated in noise \cite{langner2005improving, suni2013lombard}. Few studies have explicitly adapted Lombard speech onto speech synthesis models by focusing on articulatory effort changes \cite{raitio2011analysis, picart2014analysis}. Previously, the majority of such studies were conducted using hidden Markov model (HMM)-based statistical parametric speech synthesis (SPSS) due to its superior adaptation abilities and flexibility. The HMM model trained on normal speech was then adapted using a small amount of Lombard speech 
and improvements were shown under different noisy conditions \cite{cooke2013evaluating}. Yet, these approaches were limited to poor acoustic modeling and inability to synthesize high-fidelity speech samples. To overcome this, deep neural network approaches were implemented where the robustness of acoustic modeling is improved by efficient mapping between linguistic and acoustic features. Inspired by the success of adversarial generative models, Cycle-consistent adversarial networks (CycleGANs) showed promising results in terms of speech quality and the magnitude of the perceptual change between speech styles \cite{seshadri2019cycle, seshadri2019augmented}. An extension to recurrent neural networks and particularly long short-term memory networks (LSTMs) were proposed that it successfully adapted normal speaking style to Lombard style \cite{bollepalli2017lombard}. In \cite{bollepalli2019lombard}, the authors demonstrated results with sequence to sequence (seq2seq) TTS models along with the recently-proposed Wavenet vocoder where the audio samples are generated through a non-linear autoregressive manner. Along with different adaptation approaches, various TTS vocoders are compared in the context of style transfer and assessment was performed in terms of speaking style similarity and speech intelligibility \cite{seshadri2019vocal, bollepalli2019normal}. 

To train a TTS system with Lombard style, a sizable amount of training data is required. However, the collection of a large portion of Lombard speech is difficult. Such data sparsity limits the usage of typical data-driven approaches similar to the recent end-to-end TTS systems. Our work takes into account the use of speaking style adaptation techniques leveraging on large quantities of widely available normal speech data referred to as transfer learning. It assumes the prior knowledge from a previously model trained with large variations in linguistic and acoustic information and adapts to the target styles even with limited amount of data. In the literature, most of the vocoders for style transfer in TTS systems are either source-filter based models or convolutional models \cite{seshadri2019vocal,bollepalli2019lombard}. However, such techniques are limited by their inefficiency both in modeling proper acoustic parameters and in computational complexity of sample generation. Inspired by the performance and computational aspects of recurrent neural networks, in this work, we employ WaveRNN as a vocoder \cite{kalchbrenner2018efficient} which generates speech samples from acoustic features, i.e., mel-spectrograms. Experimental results indicate that WaveRNN is capable of adapting appropriate target speech style and able to provide more stable high-quality speech samples. To generate the mel-spectrograms from text, we utilize a popular architecture Tacotron, a seq2seq encoder–decoder neural network with attention mechanism \cite{wang2017tacotron}.

Improvement of speech intelligibility in noise can also be achieved by signal processing techniques such as amplitude compression \cite{niederjohn1976enhancement}, changes in spectral tilts \cite{lu2009contribution}, formant sharpening and dynamic range compression \cite{zorilua2014spectral}. A state-of-the-art method, referred to as Spectral Shaping and Dynamic Range Compression (SSDRC), has been shown to provide high intelligibility gains in various noisy conditions by redistributing signal energy on time-frequency information \cite{zorila2012speech}. In \cite{valentini2013combining}, the best performing method was achieved by applying additional processing, i.e., dynamic range compression after generating Lombard style adapted TTS. The results, however, failed to increase the intelligibility under competing-speaker noise. In order to develop a highly intelligible communication system and restrict the latency imposed by additional processing after the TTS synthesis, in this work, we implement Lombard-SSDRC TTS where the TTS is trained with Lombard speech processed through the SSDRC algorithm. Hence, we combine the advantages of naturally-modified Lombardness with speech enhancement strategies in frequency-domain (spectral shaping) and in time-domain (dynamic range compression) into an intelligibility-enhanced TTS synthesis system. Experimental results based on both objective and subjective evaluation confirms that the proposed method achieves remarkable performance and outplays its counterparts under both speech-shaped noise (SSN) and competing-speaker noise (CSN).

\vspace{-1mm}
\section{Text-to-Speech Synthesis}
Our proposed TTS system is composed of two separately trained neural networks: (a) Tacotron, which predicts mel-spectrograms from text and (b) WaveRNN vocoder, which converts the mel-spectrograms into time-domain waveforms.
\vspace{-2mm}
\subsection{Tacotron}

Tacotron \cite{wang2017tacotron} is a seq2seq architecture with attention mechanism and it is heavily inspired by the encoder-decoder neural network framework. The system has two main components: (a) an encoder and (b) an attention decoder. The encoder consists of 1-D convolutional filters, followed by fully-connected (FC) layers and a bidirectional gated recurrent unit (GRU). It takes text as input and extracts sequential representations of text. The attention decoder is a set of recurrent layers which produces the attention query at each decoder time-step. The input to the decoder RNN can be produced by concatenating context vector and output of the attention RNN. The decoder RNN is basically a 2-layer residual GRU whereas the attention RNN has a single GRU layer. The output of the attention decoder is a sequence of mel-spectrograms which is then passed to the vocoding stage.
\vspace{-2mm}
\subsection{WaveRNN}
\begin{figure}[t!]
  \begin{center}
    \includegraphics[height=2.5in,width=3.25in]{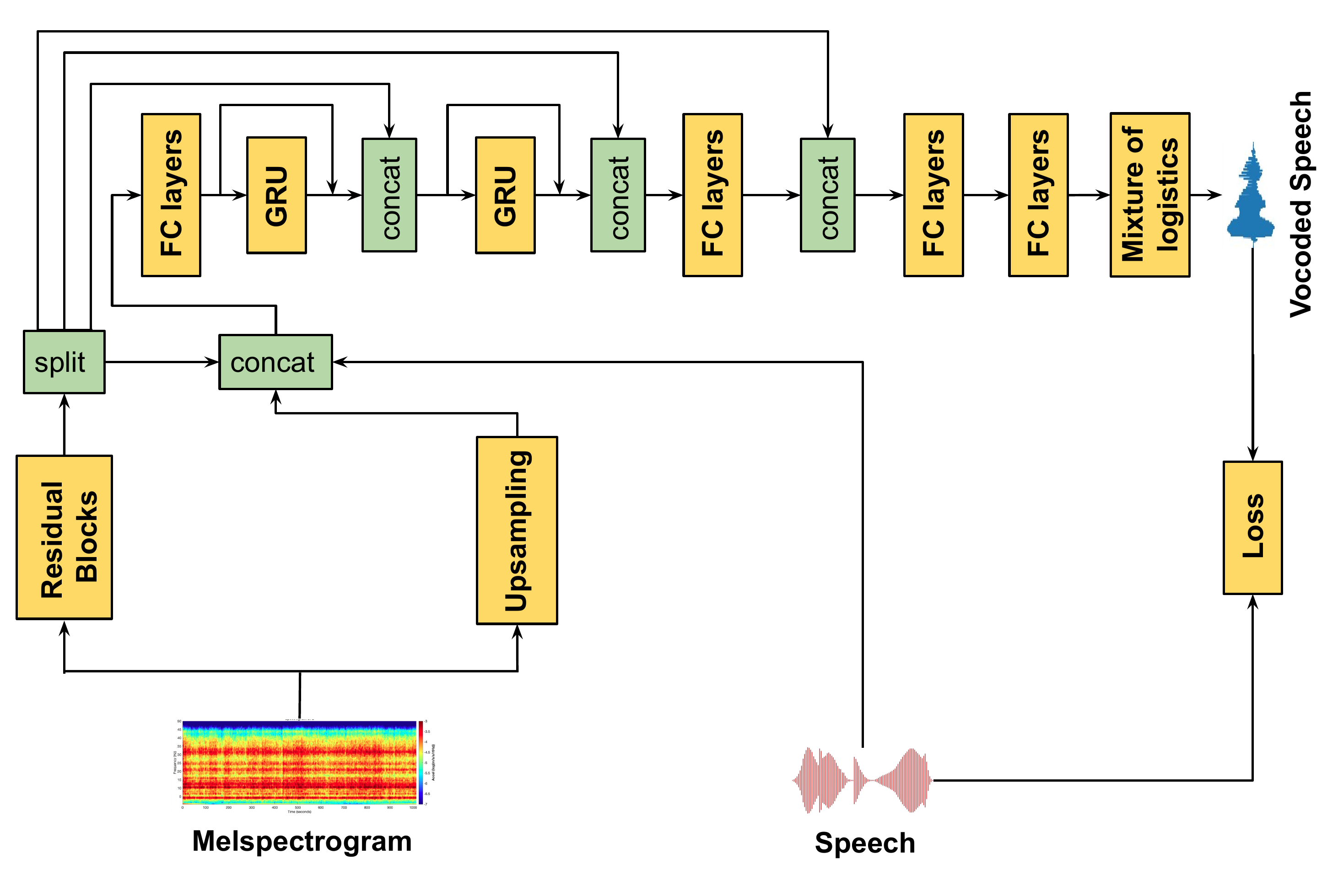}
   \vspace{-8mm}
  \end{center}
  \caption{\small Block diagram of WaveRNN architecture.}
  \vspace{-3mm}
  \label{wavernn}
\end{figure}

The implemented WaveRNN vocoder is based on the repository\footnote{\url{https://github.com/fatchord/WaveRNN}} which in turn is heavily inspired by WaveRNN training \cite{kalchbrenner2018efficient}.  This architecture is a combination of residual blocks and upsampling network, followed by GRU and FC layers as depicted in Figure \ref{wavernn}. 

The architecture can be divided into two major networks: the conditional network and the recurrent network. The conditional network consists of a pair of a residual network and an upsampling network with three scaling factors. At the input, we first map the acoustic features, i.e., the mel-spectrograms to a latent representation with the help of multiple residual blocks. The latent representation is then split into four parts which are later used as input to the subsequent recurrent network. The upsampling network is implemented to match the desired temporal size of the input signal. The outputs of these two convolutional networks i.e., residual and upsampling networks along with speech are fed into the recurrent network. As part of the recurrent network, two uni-directional GRUs are employed with a few FC layers. By designing, such network not only reduces the overhead complexity with less parameters but also it takes advantage of temporal context resulting in better prediction. 
In addition, we apply continuous univariate distribution to be a mixture of logistic distributions \cite{oord2017parallel} which allows us to easily calculate the probability on the observed discretized value.
Finally, discretized mix logistic loss is applied on the discretized speech samples.

\vspace{-1mm}
\section{Spectral Shaping and Dynamic Range Compression}
SSDRC \cite{zorila2012speech} is a signal processing approach to improve intelligibility of modifying speech when listening in noisy acoustic conditions. It comprises of a two stage process: spectral shaping followed by dynamic range compression.

\subsection{Spectral shaping}
At the SS module, the input speech is processed through three layers of filters, two of them perform probabilistic adaptive spectral sharpening. This is then followed by a fixed spectral shaping filter to boost the high frequency components. Let $x(t)$ be the input speech, Discrete Fourier Transform (DFT) is performed to obtain the magnitude spectral components $X(\omega,t)$. On the adaptive spectral shaping, the local maxima (kind of formants) are sharpened by a spectral sharpening filter $H_{s}(\omega,t)$ followed by an high frequency booster $H_{p}(\omega,t)$ . Both of the filters update their coefficients adaptively based on the voicing probability of individual frames~\cite{zorilua2014spectral}. Hence, the adaptive spectral shaped signal can be written as
\[\vspace{-1mm}
Y_{aSS}(\omega,t)=H_{s}(\omega,t)\ H_{p}(\omega,t) \ X(\omega,t).
\]
A non-adaptive pre-emphasis filter $H_{r}(\omega,t)$ is then modifying the spectra by enhancing the frequency components that are falling between 1000Hz and 4000Hz by a factor of 12dB, while reducing the energy for the frequencies below 500 Hz by 6dB/octave. The spectrally-shaped signal can be expressed as
\[\vspace{-1mm}
Y_{SS}(\omega,t)=H_{r}(\omega,t) \ Y_{aSS}(\omega,t)
\]
Inverse Fourier transform and overlap add are applied to get the spectrally-enhanced speech wavwform.

\subsection{Dynamic range compression}
DRC is a time-domain operation where the objective is to reduce the envelope variation of the speech. This has been done through modifying speech samples in each segment adaptive to the temporal envelopes. DRC is also a two step process.  In the first stage, envelope is dynamically compressed with recursive smoothing. The smoothed envelope projected onto the input output envelope characteristic (IOEC) curve gives the dynamic range compression gain. Finally, the spectral shaped output from the SS module is multiplied by the estimated gains in the DRC to provide the final intelligibility enhanced speech.


\vspace{-1mm}
\section{Transfer Learning}

The majority of deep learning methods perform well under the standard assumption that the training and inference data are drawn from similar feature space and data distribution. When the distribution changes, models need to be trained from scratch using new training data. Under the condition of data scarcity such as in our case for Lombard data, training a new model on such a limited sample size might lead to poor execution. In such cases, transfer learning (TL) offers a desirable and extremely important adaptation framework \cite{pan2009survey}. Assuming that there are two tasks, source task and target task, TL tries to 
boost the performance of the target task by utilizing knowledge learned from the source task via fine-tuning prior distributions of the hyper-parameters.

\begin{figure}[h!]
  \begin{center}
    \includegraphics[height=1.3in,width=3.1in]{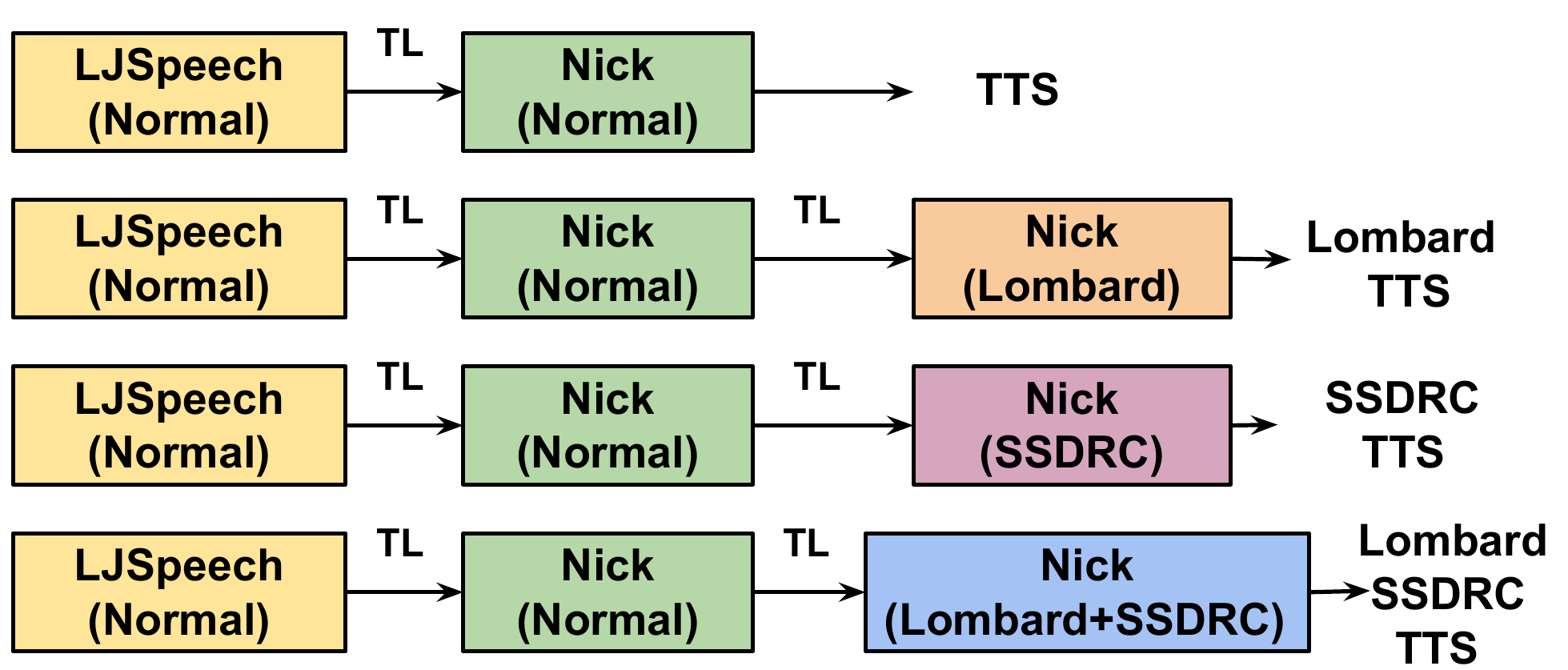}
  \end{center}
  \vspace{-3mm}
  \caption{\small A functional block diagram of the proposed adaptation techniques used in this study. Each block represents a TTS system (Tacotron + WaveRNN) which takes text as input and generates speech samples.}
  \label{tlearn}
\end{figure}

We develop four TTS systems based on the speaking styles: normal TTS, Lombard TTS, SSDRC TTS and Lombard-SSDRC TTS. To effectively transfer the prior knowledge, we initially train the TTS system with normal speech (single female speaker from LJSpeech corpora) which has a large amount of linguistic variability. Then, we adapt the learned model with normal speech from a male speaker (Nick). This normal TTS serves as the baseline system for our experiments. Lombard TTS system is then fine-tuned using again the TL approach on the limited Lombard data from the same male speaker (Nick). Whereas, SSDRC TTS uses training data processed with SSDRC algorithm applied on Nick normal speech. The last TTS system is fine-tuned on data that is prepared by applying SSDRC algorithm on Nick's Lombard speech, referred to as Lombard-SSDRC TTS. Please note that all proposed TTS systems comprise of Tacotron and WaveRNN modules \cite{paul2020} and each module is trained separately using data from the corresponding target speech style.

\vspace{-1mm}
\section{Experimental Setup}
The proposed TTS systems are trained using two publicly available database, i.e., LJSpeech corpus \cite{ljspeech17} and Nick Hurricane Challenge speech data \cite{cooke2013hurricane}. LJspeech consists of 13,100 short audio clips of a single female professional speaker reading passages. The Nick data has both normal and Lombard styles of British male voice professional speech. The normal speech consists of 2592 utterances ($\sim$2 hours) whereas the Lombard speech data has 720 utterances ($\sim$30 minutes). During training, we always consider 2400 utterances for normal and 500 utterances for Lombard speech. We additionally compare with the baseline Lombard TTS system which is built on Tacotron and WaveNet architecture \cite{bollepalli2019lombard}. The WaveNet configuration used in their system consists of three repetitions of a 10-layer convolution stack with exponentially growing dilations, 64 residual channels and 128 skip channels whereas the Tacotron architecture is similar to ours. The proposed Tacotron and WaveRNN models use 80 dimensional normalized mel-spectrograms, extracted from audio frames of width 50ms, hop length of 12.5ms and 2048-point Fourier transform. In Tacotron, character embeddings are set to 256 and a progressive training schedule is employed with reducing batch size from 32 to 8. WaveRNN architecture is based on a set of 10-layer convolution stack inside residual blocks followed by 2 GRUs. Each GRU has 512 hidden units. Code and audio samples can be found in {\footnote{\scriptsize{\url{https://dipjyoti92.github.io/TTS-Style-Transfer/}}}}. 

\vspace{-1mm}
\section{Results and Discussion}
\subsection{Objective evaluation}
In this section, objective intelligibility scores are computed and the performance of the five style adapted methods (TTS, Lombard TTS \cite{bollepalli2019lombard}, proposed Lombard TTS, also refer to as Lombard TTS (ours), SSDRC TTS and Lombard-SSDRC TTS) under two different noisy conditions are compared. A recently developed  intelligibility metric called `speech intelligibility in bits' ($SIIB^{Gauss}$) \cite{van2018evaluation} is implemented as an objective evaluation metric. It takes into account the information capacity of a Gaussian channel between clean and noisy signals. Higher values refer to better intelligibility. The scores are evaluated from 250 utterances and each adaptation approach has 50 distinct utterances. Table \ref{siigauss} presents $SIIB^{Gauss}$ intelligibility scores. We consider three different Signal-to-Noise Ratio (SNR) levels masked with two types of noise: speech-shaped noise (0 , -5 and -10 dB) and competing-speaker (-7, -14 and -21 dB). Since we are focusing in the context of TTS, we omitted the scores for natural speech in our experiments.
\begin{table}[t!]
\footnotesize
\renewcommand{\arraystretch}{1.2}
\setlength{\tabcolsep}{2.2pt}
\caption{$SIIB^{Gauss}$ intelligibility measure at different SNR levels under 
speech-shaped and competing-speaker noise.}
\vspace{-1mm}
\centering
\begin{tabular}{cccclccc}
\specialrule{1.25pt}{1pt}{1pt}
Systems & \multicolumn{3}{c}{SSN}                    &  & \multicolumn{3}{c}{CSN}                     \\ \cline{2-4} \cline{6-8} 
                    & -10 dB          & -5 dB          & 0 dB          &  & -21 dB         & -14 dB         & -7 dB          \\ \specialrule{1.25pt}{1pt}{1pt}
TTS                 & 15.03         & 26.80         & 42.43          &  & 13.3           & 17.86          & 28.27          \\
Lombard TTS {\cite{bollepalli2019lombard}} & 17.89          & 33.89          & 54.53          &  & 9.91           & 18.1           & 36.21          \\
Lombard TTS (ours)  & 20.02          & 37.43          & 58.65          &  & 13.52          & 22.51          & 41.65          \\
SSDRC TTS           & 29.90          & 51.02         & 77.97          &  & 16.73          & 29.75          & 55.56          \\
Lombard-SSDRC TTS   & \textbf{35.04} & \textbf{58.68} & \textbf{88.35} &  & \textbf{19.13} & \textbf{35.84} & \textbf{68.35} \\ \specialrule{1.25pt}{1pt}{1pt}
\end{tabular}
\vspace{-4mm}
\label{siigauss}
\end{table}

It can be observed that the standard synthesis system trained with normal speech, referred to here as the speech type ‘TTS’, is the worst performer when compared to the rest of the methods under any condition as expected. To enhance the intelligibility, TTS is re-trained with limited Lombard style data. We observe that the proposed Lombard TTS i.e., Lombard TTS (ours) is able to successfully mimic the Lombardness and outperforms baseline Lombard TTS from \cite{bollepalli2019lombard} with a relative improvement between 8\% and 12\% in SSN and 15\% to 36\% in CSN conditions across different SNR levels: from low to high SNRs. The results also show high performance gain of 18\% and 36\% in Low SNR i.e., -10 dB for SSN and -21 dB in CSN conditions, respectively. The use of WaveRNN instead of WaveNet vocoder as in the baseline Lombard TTS, demonstrates how the choice of vocoder  affects the intelligibility of synthesized speech. WaveRNN effectively adapts to the new style while trained with limited amount of target style data. Furthermore, taking into account the SSDRC approach, we aim towards additional intelligibility gains under adverse noise conditions. Our results reveal that SSDRC TTS archives further improvement compared to the Lombard TTS. Motivating by the boosting effect of Lombard style, along with the enhancement by SSDRC data in terms of speech intelligibility, the proposed Lombard-SSDRC TTS shows significant intelligibility gains between 110\% and 130\% in SSN, and 47\% to 140\% in CSN against TTS. Those results can be attributed by the fact that the combined model exploits efficiently both Lombardness and spectral shaping with range compression by modifying time-frequency regions.
\vspace{-3mm}
\begin{figure}[t!]
\centering
    \includegraphics[height=1.8in, width=2.8in]{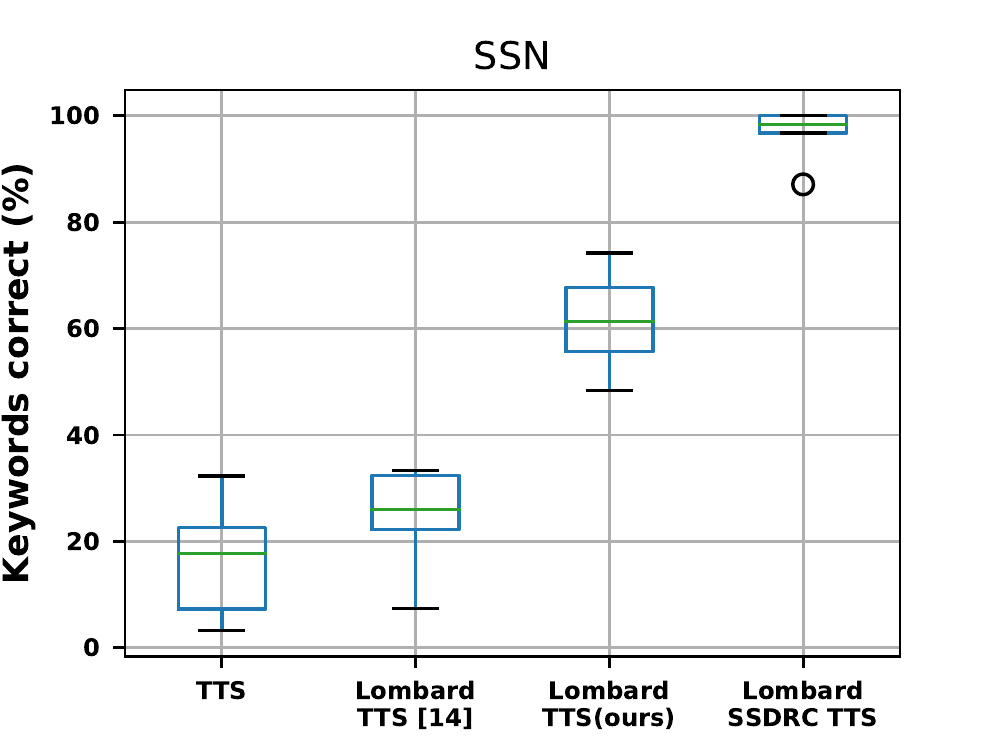}
    \vspace{-3mm}
    \includegraphics[height=1.8in, width=2.8in]{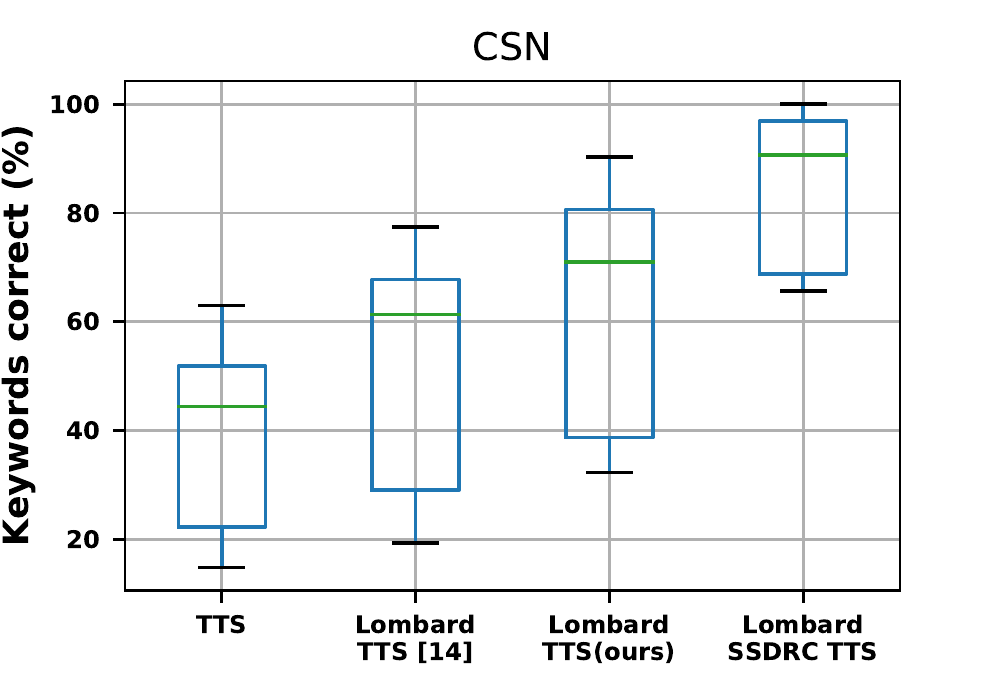}
  \caption{\small Box plot results for listeners’ keyword scores across of methods for SSN and CSN.}
  \label{subjective}
  \vspace{-7mm}
\end{figure}

\subsection{Subjective evaluation}
To assess the performance on subjective evaluation, metric scores were computed based on the number of keywords correctly identified in each sentence. The short common words ‘a’, ‘the’, ‘in’, ‘to’, ‘on’, ‘of’, and ‘for’ were excluded. The listening test was conducted via a web-based interface and ten native listeners participated in the test. No listener heard the same sentence twice, and each condition was heard by the same number of listeners. Since intelligibility level varies from one listener to another and large variability in scores can be possible when listeners use different hearing devices or backgrounds, intelligibility gains should be observed from a common reference point. This was achieved by designing an initial pilot study where subject-specific SNR levels are matched with the speech reception threshold (SRT) at which 40\% of normal speech is intelligible for each individual listener. In the final listening test, we choose SNR levels based on the values obtained from the pilot study for each listener individually.

Box plots reported in Figure \ref{subjective} allow comparison between different TTS modification algorithms. The subjective results reveal a similar pattern to the objective metrics. The proposed Lombard-SSDRC TTS outperforms all other methods with a remarkable margin under all noisy conditions. Lombard-SSDRC TTS shows superior performance by achieving a remarkable relative improvement of 455\% for SSN and 104\% for CSN in median keyword correction rate compared to TTS method. It is worth noting that the performance gains are immensely higher in SSN condition, although we observe outstanding performance gains in both noisy conditions. Moreover, the comparison between Lombard TTS \cite{bollepalli2019lombard} and Lombard TTS (ours) adaptation methods highlights that Lombard TTS (ours) method achieves significantly better performance in terms of keyword correction rate. This confirms the adaptability of WaveRNN for limited data scenarios, and shows its effectiveness in the transfer learning approach. The results indicate a relative improvement of 136\% in SSN and 16\% in CSN compared to Lombard TTS \cite{bollepalli2019lombard} in terms of median keyword correction rate.

\vspace{-1mm}
\section{Conclusion}

In this paper, we performed transfer learning and constructed adapted Tacotron+WaveRNN TTS systems for speaking style modification. The synthesized speech was modified with two strategies: Lombard style recordings and SSDRC algorithm. First, we showed that the Lombard-adapted TTS system (ours) is able to successfully learn Lombard style under limited training data and outperforms the baseline Lombard TTS system \cite{bollepalli2019lombard} by a significant margin when masked either with SSN or CSN noise. This shows the advantage of applying neural-based WaveRNN vocoder and its importance in achieving highly-intelligible Lombard synthetic speech. Furthermore, to enjoy larger intelligibility gains, we combined the benefits of Lombardness with the SSDRC modification strategy. Our experiments on both objective and subjective intelligibility scores confirmed that both modifications contributed to significant gains under all noisy conditions.
In the future, we would like to investigate whether similar intelligibility gains can be obtained by applying cross-speaker adaptation.

{\bf Acknowledgements:} The work has received funding from the EUs H2020 research and innovation programme under the MSCA GA 67532 (the ENRICH network: www.enrich-etn.eu).

\bibliographystyle{IEEEtran}
\bibliography{mybib}

\begin{thebibliography}{10}
\providecommand{\url}[1]{#1}
\csname url@samestyle\endcsname
\providecommand{\newblock}{\relax}
\providecommand{\bibinfo}[2]{#2}
\providecommand{\BIBentrySTDinterwordspacing}{\spaceskip=0pt\relax}
\providecommand{\BIBentryALTinterwordstretchfactor}{4}
\providecommand{\BIBentryALTinterwordspacing}{\spaceskip=\fontdimen2\font plus
\BIBentryALTinterwordstretchfactor\fontdimen3\font minus
  \fontdimen4\font\relax}
\providecommand{\BIBforeignlanguage}[2]{{%
\expandafter\ifx\csname l@#1\endcsname\relax
\typeout{** WARNING: IEEEtran.bst: No hyphenation pattern has been}%
\typeout{** loaded for the language `#1'. Using the pattern for}%
\typeout{** the default language instead.}%
\else
\language=\csname l@#1\endcsname
\fi
#2}}
\providecommand{\BIBdecl}{\relax}
\BIBdecl

\bibitem{lu2008speech}
Y.~Lu and M.~Cooke, ``Speech production modifications produced by competing
  talkers, babble, and stationary noise,'' \emph{The Journal of the Acoustical
  Society of America}, vol. 124, no.~5, pp. 3261--3275, 2008.

\bibitem{hansen1996analysis}
J.~H. Hansen, ``Analysis and compensation of speech under stress and noise for
  environmental robustness in speech recognition,'' \emph{Speech
  Communication}, vol.~20, no. 1-2, pp. 151--173, 1996.

\bibitem{cooke2013evaluating}
M.~Cooke, C.~Mayo, C.~Valentini-Botinhao, Y.~Stylianou, B.~Sauert, and Y.~Tang,
  ``Evaluating the intelligibility benefit of speech modifications in known
  noise conditions,'' \emph{Speech Communication}, vol.~55, no.~4, pp.
  572--585, 2013.

\bibitem{valentini2014intelligibility}
C.~Valentini-Botinhao, J.~Yamagishi, S.~King, and R.~Maia, ``Intelligibility
  enhancement of hmm-generated speech in additive noise by modifying mel
  cepstral coefficients to increase the glimpse proportion,'' \emph{Computer
  Speech \& Language}, vol.~28, no.~2, pp. 665--686, 2014.

\bibitem{erro2014enhancing}
D.~Erro, T.~C. Zoril{\u{a}}, and Y.~Stylianou, ``Enhancing the intelligibility
  of statistically generated synthetic speech by means of noise-independent
  modifications,'' \emph{IEEE/ACM Transactions on Audio, Speech, and Language
  Processing}, vol.~22, no.~12, pp. 2101--2111, 2014.

\bibitem{valentini2013combining}
C.~Valentini-Botinhao, J.~Yamagishi, S.~King, and Y.~Stylianou, ``Combining
  perceptually-motivated spectral shaping with loudness and duration
  modification for intelligibility enhancement of hmm-based synthetic speech in
  noise.'' in \emph{Proc. Interspeech}, 2013, pp. 3567--3571.

\bibitem{langner2005improving}
B.~Langner and A.~W. Black, ``Improving the understandability of speech
  synthesis by modeling speech in noise,'' in \emph{Proc. IEEE International
  Conference on Acoustics, Speech, and Signal Processing.}, 2005, pp. 261--265.

\bibitem{suni2013lombard}
A.~Suni, R.~Karhila, T.~Raitio, M.~Kurimo, M.~Vainio, and P.~Alku, ``Lombard
  modified text-to-speech synthesis for improved intelligibility: submission
  for the hurricane challenge 2013.'' in \emph{Proc. Interspeech}, 2013, pp.
  3562--3566.

\bibitem{raitio2011analysis}
T.~Raitio, A.~Suni, M.~Vainio, and P.~Alku, ``Analysis of {HMM}-based {L}ombard
  speech synthesis,'' in \emph{Proc. Interspeech}, 2011.

\bibitem{picart2014analysis}
B.~Picart, T.~Drugman, and T.~Dutoit, ``Analysis and {HMM}-based synthesis of
  hypo and hyperarticulated speech,'' \emph{Computer Speech \& Language},
  vol.~28, no.~2, pp. 687--707, 2014.

\bibitem{seshadri2019cycle}
S.~Seshadri, L.~Juvela, J.~Yamagishi, O.~R{\"a}s{\"a}nen, and P.~Alku,
  ``Cycle-consistent adversarial networks for non-parallel vocal effort based
  speaking style conversion,'' in \emph{Proc. IEEE International Conference on
  Acoustics, Speech and Signal Processing (ICASSP)}, 2019, pp. 6835--6839.

\bibitem{seshadri2019augmented}
S.~Seshadri, L.~Juvela, P.~Alku, O.~R{\"a}s{\"a}nen \emph{et~al.}, ``Augmented
  {C}ycle{GAN}s for continuous scale normal-to-lombard speaking style
  conversion,'' \emph{Proc. Interspeech 2019}, pp. 2838--2842, 2019.

\bibitem{bollepalli2017lombard}
B.~Bollepalli, M.~Airaksinen, and P.~Alku, ``Lombard speech synthesis using
  long short-term memory recurrent neural networks,'' in \emph{Proc. IEEE
  International Conference on Acoustics, Speech and Signal Processing}, 2017,
  pp. 5505--5509.

\bibitem{bollepalli2019lombard}
B.~Bollepalli, L.~Juvela, P.~Alku \emph{et~al.}, ``Lombard speech synthesis
  using transfer learning in a {T}acotron text-to-speech system,'' \emph{in
  Proc. Interspeech}, pp. 2833--2837, 2019.

\bibitem{seshadri2019vocal}
S.~Seshadri, L.~Juvela, O.~R{\"a}s{\"a}nen, and P.~Alku, ``Vocal effort based
  speaking style conversion using vocoder features and parallel learning,''
  \emph{IEEE Access}, vol.~7, pp. 17\,230--17\,246, 2019.

\bibitem{bollepalli2019normal}
B.~Bollepalli, L.~Juvela, M.~Airaksinen, C.~Valentini-Botinhao, and P.~Alku,
  ``Normal-to-{L}ombard adaptation of speech synthesis using long short-term
  memory recurrent neural networks,'' \emph{Speech Communication}, vol. 110,
  pp. 64--75, 2019.

\bibitem{kalchbrenner2018efficient}
N.~Kalchbrenner, E.~Elsen, K.~Simonyan, S.~Noury, N.~Casagrande, E.~Lockhart,
  F.~Stimberg, A.~Oord, S.~Dieleman, and K.~Kavukcuoglu, ``Efficient neural
  audio synthesis,'' in \emph{International Conference on Machine Learning},
  2018, pp. 2410--2419.

\bibitem{wang2017tacotron}
Y.~Wang, R.~Skerry-Ryan, D.~Stanton, Y.~Wu, R.~J. Weiss, N.~Jaitly, Z.~Yang,
  Y.~Xiao, Z.~Chen, S.~Bengio \emph{et~al.}, ``Tacotron: Towards end-to-end
  speech synthesis,'' \emph{arXiv preprint:1703.10135}, 2017.

\bibitem{niederjohn1976enhancement}
R.~Niederjohn and J.~Grotelueschen, ``The enhancement of speech intelligibility
  in high noise levels by high-pass filtering followed by rapid amplitude
  compression,'' \emph{IEEE Transactions on Acoustics, Speech, and Signal
  Processing}, vol.~24, no.~4, pp. 277--282, 1976.

\bibitem{lu2009contribution}
Y.~Lu and M.~Cooke, ``The contribution of changes in f0 and spectral tilt to
  increased intelligibility of speech produced in noise,'' \emph{Speech
  Communication}, vol.~51, no.~12, pp. 1253--1262, 2009.

\bibitem{zorilua2014spectral}
T.~C. Zoril{\u{a}} and Y.~Stylianou, ``On spectral and time domain energy
  reallocation for speech-in-noise intelligibility enhancement,'' in
  \emph{Proc. Interspeech}, 2014.

\bibitem{zorila2012speech}
T.~C. Zorila, V.~Kandia, and Y.~Stylianou, ``Speech-in-noise intelligibility
  improvement based on spectral shaping and dynamic range compression,'' in
  \emph{Proc. Interspeech}, 2012.

\bibitem{oord2017parallel}
A.~Oord, Y.~Li, I.~Babuschkin, K.~Simonyan, O.~Vinyals, K.~Kavukcuoglu,
  G.~Driessche, E.~Lockhart, L.~Cobo, F.~Stimberg \emph{et~al.}, ``Parallel
  {W}ave{N}et: Fast high-fidelity speech synthesis,'' in \emph{International
  Conference on Machine Learning}, 2018, pp. 3918--3926.

\bibitem{pan2009survey}
S.~J. Pan and Q.~Yang, ``A survey on transfer learning,'' \emph{IEEE
  Transactions on Knowledge and Data Engineering}, vol.~22, no.~10, pp.
  1345--1359, 2009.

\bibitem{paul2020}
D.~Paul, Y.~Pantazis, and Y.~Stylianou, ``Speaker conditional wavernn: Towards
  universal neural vocoder for unseen speaker and recording conditions,''
  \emph{arXiv:2008.05289}, 2020.

\bibitem{ljspeech17}
Keithito, ``The {LJ}speech dataset,''
  \url{https://keithito.com/LJ-Speech-Dataset/}, 2017.

\bibitem{cooke2013hurricane}
M.~Cooke, C.~Mayo, C.~Valentini-Botinhao \emph{et~al.}, ``Hurricane natural
  speech corpus,'' \emph{LISTA Consortium, Language and Speech Laboratory,
  Universidad del Pais.}, 2013.

\bibitem{van2018evaluation}
S.~Van~Kuyk, W.~B. Kleijn, and R.~C. Hendriks, ``An evaluation of intrusive
  instrumental intelligibility metrics,'' \emph{IEEE/ACM Transactions on Audio,
  Speech, and Language Processing}, vol.~26, no.~11, pp. 2153--2166, 2018.

\end{thebibliography}


\end{document}